\documentclass[aps,prb,amsmath,amssymb,amsfonts,twocolumn,nofootinbib]{revtex4}
\usepackage{graphicx}
\usepackage{dcolumn}
\usepackage{bm}
\usepackage{amsmath}
\usepackage{amssymb}
\usepackage{color}

\usepackage{soul}

\makeatletter
\renewcommand*\env@matrix[1][*\c@MaxMatrixCols c]{%
  \hskip -\arraycolsep
  \let\@ifnextchar\new@ifnextchar
  \array{#1}}
\makeatother

\newcommand{\be}{\begin{equation}}
\newcommand{\ee}{\end{equation}}

\newcommand{\bea}{\begin{eqnarray}}
\newcommand{\eea}{\end{eqnarray}}
\newcommand{\la}{\langle}
\newcommand{\ra}{\rangle}
\newcommand{\dg}{\dagger}
\newcommand{\td}{\tilde}
\newcommand{\dgc}{\maltese}

\begin{document}

\title{Lindblad equation for a non-interacting fermionic system: full-counting statistics}

\author{M. V. Medvedyeva}
\affiliation{Department of Physics, Georg-August-Universitaet Goettingen,
Friedrich-Hund-Platz 1, 37077 Goettingen, Germany}
\author{S. Kehrein}
\affiliation{Department of Physics, Georg-August-Universitaet Goettingen,
Friedrich-Hund-Platz 1, 37077 Goettingen, Germany~}

\begin{abstract}
We develop a method of calculating the full-counting statistics for a non-interacting fermionic system coupled to the memory-less reservoirs. 
The evolution of the system is described by the Lindblad equation. 
By the basis change the Liouvillian operator is brought to the quadratic form. This allows us a straightforward calculation of any observable in the non-equilibrium steady state.
We introduce the counting field in the Lindblad equation which brings us to the generating function and helps us to obtain all cumulants of the charge transport. 
For the two-site system we give the expression for the generating function. 
For system  longer than two sites we perform numerical investigations which suggest that {\it in a uniform system the cumulants of order $k$ are independent of the size of the system for system sizes larger $k+1$.} 
The counting statistics from the Lindblad approach does not take into account interference in the reservoirs which gives a decreased noise in comparison with the Green function method which describes phase coherent leads. The current obtained by two methods is the same, which relies on the current conservation.  
The Fano factors are different (with a linear relation connecting them) and allow to distinguish between memory-less and phase coherent reservoirs. 
\end{abstract}

\maketitle

\section{Introduction}

The influence of the environment can significantly change the 
behavior of a quantum system. 
For example, the resonant level model experiences a phase transition when 
coupled to the interacting bath, which was found both analytically and experimentally~\cite{RNM}. 

The simplest way to describe an open quantum system is to write down a linear differential equation for the
evolution of the density matrix, which obey three main properties:  the trace 
of density matrix, its hermicity and its positive-semidefinitness are preserved. This class of equations is called Lindblad equations~\cite{lindblad, breuer}.
The consequence of such formulation of the interaction between the system and the bath implies the absence of  memory. 
The Lindblad equation can describe sufficiently well
the transport through a molecular system coupled to the leads~\cite{molec}, 
decay of the atoms due to irradiation~\cite{kleinert}, for example in optical lattices~\cite{bonn}, 
electronic transport through a system in the Coulomb blockade~\cite{Naz, Count}. 

In this work we study the transport properties through a fermionic non-interacting system coupled to non-interacting leads. 
The problem can be formulated in the language of the Lindblad equation under the assumption of infinite bias voltage in the leads~\cite{Gurv}. 
For infinite bias voltage, the particles in the reservoirs are much faster then in the system. Therefore, after electron hops from the system to the reservoir it escapes infinitely fast from the contact and mixes with other electrons; and when electron hops from the reservoir to the system, the hole in its place disappears infinitely fast. The process of the electron hopping into and out of the system is hence a classical process without memory. 
The full-counting statistics in the regime of the strong Coulomb blockade has been recently investigated in Ref.~\onlinecite{Nov} also for the non-Markovian bath. 

So far two almost equivalent approaches have been developed to consider the Lindblad equation for non-interacting fermionic systems. One approach is based on the formulation on the language of Majorana fermions~\cite{Prosen10,Prosen12}. The idea of the other approach is introducing the Hilbert space which combines bra and ket vectors and allows us to represent the density matrix as a vector in this Hilbert space~\cite{Kosov}. 

In our work we follow the method of Ref.~\onlinecite{Kosov}. We extend this method to include the full-counting statistics~\cite{prepare}.
We analyze the current and the noise in a uniform 1d wire. The current though the system is independent of the length of the system. 
It is in accordance with physical intuition: the electrons come in the system, propagate freely and escape, and the region of the free propagation
does not influence the current. This result is also in a full agreement with Meir-Wingreen formula~\cite{MW} under the assumption of the infinite bias voltage in the leads. The result for the noise in the Lindblad formalism is different from the one obtained by the Meir-Wingreen type of the approach. 
We ascribe it to different types of measurements of the current-noise statistics for the Lindblad and Meir-Wingreen formulations.

\section{Lindblad equation for a non-interacting fermionic system}
In this section we review the super-operator formalism 
for describing the evolution of the density matrix of an open quantum system of non-interacting fermions, which leads to an easy way of computing the observables in the non-equilibrium steady state (NESS)~\cite{Kosov}. 
Then we explicitly show how to build a new convenient operator basis for a tight-binding chain linearly coupled to the bath. This allows us to calculate any observable in NESS, for example the current. 
At the end we introduce a counting field in the Lindblad equation~\cite{Groth} and develop a way of getting a full-counting statistics in super-fermion formulation of the Lindblad equation.
We derive the generating function for the two-site system and analyze the statistics of the charge transport through this simple system. 

We consider a fermionic system coupled to the reservoirs at infinite chemical potential. 
Under this assumption the dynamics in the system can be described by the Lindblad equation~\cite{Gurv}:
\bea i\frac{d\rho}{dt}&=&\mathcal{L} \rho, \label{define}\\
 \mathcal{L}\rho &=& [H,\rho] + i \sum_{\mu}\left(2 L_\mu  \rho L_\mu^\dg - \{L_\mu L_\mu^\dg,\rho\} \right). \label{liouv}\eea
Operator $\mathcal{L}$ is called the Liouvillian, $H$ is the Hamiltonian,  $L_\mu$ are the Lindblad operators describing the coupling to the bath. 
The coupling to the reservoirs allows for hopping into or out of the system: 
\be L_{\mu}=\sqrt{\Gamma^{out}_\mu} a_\mu, L_{\mu}=\sqrt{\Gamma^{in}_\mu} a^\dg_\mu \label{lindblad}, \ee
where $\Gamma$ is a hybridization between the system and the bath. For simplicity we assume the hybridization to be local in space.
The system itself consists of non-interacting fermions described by the tight-binding Hamiltonian:
\be H=  \sum_{\{ij\}} t_{ij} \left( a^\dg_ia_j+ h.c \right)+ \sum_i U_i a^\dg_ia_i ,\label{ham}\ee
where $\{ij\}$ is responsible for the shape of the lattice,  $t_{ij}$ is the hopping strength between sites $i$ and $j$, $U_i$ is an on-site potential.   

\subsection{Operators and tilde operators}
\label{subsec:tilde}
Here we introduce the operators and tilde operators which act on bra and ket vectors correspondingly. We follow the Ref.~\onlinecite{Kosov}.

In the second quantization excitations of the system are described by the set of the creation and annihilation operators $\{a\}, \{a^\dg\}$. 
In the Liouvillian one encounters terms of the form:
\be \mathcal{L}\rho: f(a,a^\dagger) \rho, \rho f(a,a^\dagger), f(a,a^\dagger) \rho g(a,a^\dagger), \label{l1}\ee
where $f$ and $g$ are arbitrary functions of the operators $a,a^\dagger$.
Let us represent the operators which are acting from the left as a separate set of operators. We denote them by $\{\td{a}\}$, $\{\td{a}^\dg\}$.  
Then we can write terms from (\ref{l1}) as
\be \mathcal{L}\rho: f(a,a^\dagger) \rho, \td{f}(\td{a}, \td{a}^\dagger) \rho , f(a,a^\dagger) \td{g}(\td{a}, \td{a}^\dagger)\rho. \label{l2}\ee
Now all operators act on the density matrix from the left. 

Let us introduce the Fock space for such representation of the Liouvillian. It is the cross product of two identical Fock spaces of the states of the system $|m)$:
\be |mn\ra =|m\ra  \times |n\ra \ee
where the first entry corresponds to the state  $|m)$  and  the second to $(n|$  in the conventional wave-function Hilbert space. Operators $a$ and $a^\dg$ act on $|m\ra$ while the tilde operators $\td{a}$ and $\td{a}^\dg$ act on $|n\rangle$.

The action of the tilde-operators, $\td{a}$ and $\td{a}^\dg$, is specified in the following way:
\bea \td{a} |m n\rangle &=& i (-1)^{m+n} |m) ( n | a^{\dg}, \nonumber \\
\td{a}^{\dg} |m n\rangle &=& i (-1)^{m+n} |m) ( n | a  \nonumber.\eea

The operators $a$, $a^\dg$ and $\td{a}$, $\td{a}^\dg$ anticommute
\be \{a_i,a_j^\dg \}=\delta_{ij}, \{\td{a}_i,\td{a}_j^\dg \}=\delta_{ij},\ee
and different "flavors" also mutually anticommute:
\be \{a_i,\td{a}_j^\dg \}=0, \{ a_i,\td{a}_j \}=0,
\{a_i,a_j \}=0, \{\td{a}_i,\td{a}_j \}=0.
\ee

Any operator with matrix elements $A_{mn}=( n| A | m )$ is represented in the Liouville-Fock space as
\be |A \ra = \sum_{mn} A_{mn} |mn\ra.\ee
In particular, we can determine the identity operator
\be |I \ra = \sum_{n} (-i)^{\sum n_i} |nn\ra,\ee
where we write for shortness the element of the many particle basis $|n_1 n_2 \ldots n_N\ra$ as $|n\ra$.
The expectation value of any operator $A$ is computed as
\be Tr(A \rho) = \la I | A\rho\ra,\ee
where $\la I |$ is a left-vacuum of the Liouvillian.

\subsection{New operator basis}
\label{subsec:newbasis}

Let us rewrite the Liouvillian~(\ref{liouv})  for the non-interacting fermions on the lattice with the Hamiltonian~(\ref{ham}) coupled to the environment~(\ref{lindblad}) using operators and tilde-operators:
\begin{widetext}
\bea\mathcal{L} = t \sum_{\{ij\}} \left[\left( a^\dg_ia_j+ h.c \right)-\left( \td{a}^\dg_i \td{a}_j+ h.c \right)\right] -2\sum_{\mu} \Gamma_\mu^{out}\td{a_\mu}a_\mu -2\sum_{\mu} \Gamma_\mu^{in}\td{a_\mu}^\dg a_\mu^\dg -\nonumber\\
- i\sum_\mu \Gamma_\mu^{out}(a_\mu^\dg a_\mu + \td{a_\mu}^\dg \td{a_\mu})+
i\sum_\mu \Gamma_\mu^{in}(a_\mu^\dg a_\mu + \td{a_\mu}^\dg \td{a_\mu})-2i\sum_\mu \Gamma_\mu^{out}.\eea
We can diagonalize $\mathcal{L}$ by the particle-hole transformation
$ \td{a}=b^\dg, \td{a}^\dg=b$. Then the Liouvillian can be expressed as
\be \label{Ldiagonal}
\mathcal{L} = (a^\dg b^\dg) M \begin{pmatrix}
                               a \\
                               b
                              \end{pmatrix}
 -i\sum_\mu \Gamma_\mu^{out} - i\sum_\mu \Gamma_\mu^{in}.\ee
$M$ is a matrix which takes into account the tight-binding structure of the Liouvillian and the coupling to the environment. 

As an example we write down the matrix $M$ for a one-dimensional chain with constant hopping and constant on-site energy. The electrons can only hop into the chain at the first site, and hop out of the chain at the last site (there are only two Lindblad operators, $\sqrt{\Gamma_1} a_1$ and  $\sqrt{\Gamma_2} a_N^\dg$):
\be \label{m} M_{1d} = \begin{pmatrix}[ccccc|ccccc]
                        -i\Gamma_1+U & t & 0 & \ldots& 0 &0 &\ldots&\ldots&\ldots& 0 \cr
                        t & U & t & \ddots & \vdots & 0  & \ldots&\ldots&\ldots& 0 \cr
                        0 & \ddots & \ddots & \ddots & \vdots & 0& \ldots& \ldots&\ldots& 0 \cr                
                        \vdots &  \ldots &\ddots & U & t & 0 & \ldots & \ldots&\ldots& 0\cr
                        
                        0 & \ldots & 0 & t & i\Gamma_2+U & 0 & \ldots & \ldots& 0& 2\Gamma_2 \cr
                        \hline
                        -2\Gamma_1 &0 &\ldots&\ldots& 0& i\Gamma_1+U & t & 0 & \ldots& 0  \cr
                        0 &\ldots&\ldots&\ldots& 0& t & U & t & \ddots & \vdots  \cr
                        0 &\ldots&\ldots&\ldots& 0 & 0 & \ddots & \ddots & \ddots & \vdots \cr                
                        0 &\ldots&\ldots&\ldots& 0 & \vdots & \ldots & \ddots & U & t\cr
                        
                         0 &\ldots&\ldots&\ldots& 0& 0 & \ldots & 0 & t & -i\Gamma_2+U  
                       \end{pmatrix}
\ee
\end{widetext}
For the case of the coupling to the environment at arbitrary sites~(\ref{lindblad}), we get 
on the upper left diagonal terms $-i \Gamma_\mu^{out}+i \Gamma_\mu^{in}$ at positions $(\mu,\mu)$,
while on the bottom right diagonal we get the term $i \Gamma_\mu^{out}-i \Gamma_\mu^{in}$ at positions $(\mu+n,\mu+n)$.
On the diagonal of the bottom-left sub-matrix we have $-2\Gamma_\mu^{out}$ at positions $(\mu+n,\mu)$ and 
on the diagonal of the top-right submatrix we have $2\Gamma_\mu^{in}$ at positions $(\mu,\mu+n)$.

For tight-binding models of fermions the matrix $M$ is diagonalizable: 
$$D=P^{-1}MP,$$
where $P$ is the matrix consisting of the eigenvectors of $M$ and $D$ is the diagonal matrix of eigenvalues of $M$. 
Let us order the eigenvalues in the following way: first $n$ eigenvalues with the negative imaginary part, and the rest are with the positive imaginary part, such that the $i$th and $(i+n)$th eigenvalues are complex conjugate to each other (the characteristic polynomial of the matrix $M$ is real, the roots of the real polynomial are either real or come in the  pairs of complex conjugate numbers, but $M$ does not have real eigenvalues, hence all its eigenvalues come in complex conjugate pairs). 

The creation-annihilation operators in the new basis are
\be \label{gtransform} g^\dgc=(a^\dg b^\dg)P, g=P^{-1}\begin{pmatrix}
                                  a \\ b
                                 \end{pmatrix}.
\ee
They obey the anticommutation relations:
\be \{g_k,g_j^\dgc\}=\delta_{kj}, \{g_k,g_j\}=0, \{g_k^\dgc,g_j^\dgc\}=0,\label{antig}\ee
although they are not conjugate operators:
\be g^\dgc \ne (g)^\dg, \nonumber\ee
which is why we call them $ g^\dgc$ and not $g^\dg$.

The Liouvillian {\it should} have at least one zero eigenvalue, as zero eigenvalue corresponds to NESS. 
Its right eigenvector corresponds to the density matrix of NESS, $\rho_{\infty}$.   
Then there should exist a set of eigenvalues $\{\lambda_i\}$ which sum to the constant term of the Liouvillian written in the basis $g$~(\ref{Ldiagonal}):
\be \sum_{ \{\lambda_i\} } \lambda_i -i\Gamma_1-i\Gamma_2=0.\label{sumlambda}\ee
This set is formed by the eigenvalues which have positive imaginary part.
The structure of the matrix $M$ is such that its eigenvectors can be represented as $\rho_1 e^{i\phi_1},\rho_2 e^{i\phi_2}, \ldots, \rho_n, -i\rho_1 e^{i\phi_1}, -i\rho_2 e^{i\phi_2}, \ldots, -i\rho_n$. Therefore, the problem of calculating the eigenvalues of $M$ can be simplified to finding the eigenvalues of the matrix $N$ which is twice smaller than $M$. The matrix $N$ has $-i\Gamma_1$ and $-i\Gamma_2$ as the first and the last element on the main diagonal, while the elements right above and right below the main diagonal are all equal $t$.
All eigenvalues of $N$ have negative imaginary part and their sum is equal to the trace of $N$, $-i(\Gamma_1+i\Gamma_2)$, hence the sum of the complex conjugate set of eigenvalues equals  $i(\Gamma_1+\Gamma_2)$.

Using the anticommutation relation~(\ref{antig}) for the $g_i$ and $g_i^\dg$, $i=1,\ldots,n$ 
and the subsequent particle-hole transformation
\bea f_i&=&g_i, f_i^\dgc=g_i^\dgc, \label{ftransform} \nonumber \\
\td{f_i}&=&g^\dgc_{i+n}, \td{f}_i^\dgc=g_{i+n}, i=1,\ldots,n \label{ftransform2} \nonumber\eea
we bring the  Liouvillian to the diagonal form: 
\be \mathcal{L} = \sum_{i=1,\ldots,N} \lambda_i f^\dgc f - \sum_{i=1,\ldots,N} \lambda^{*}_i \td{f}^\dgc \td{f}. \label{Ldiag}\ee
Therefore, in the diagonal basis spanned by operators $f$ the NESS density matrix is the vacuum of the annihilation, operators $\{f\}$ and $\{\td{f}\}$
\be \rho_{\infty}=\beta|0\ldots0\ra_{f}, \label{rhoinf}\ee
as index in the bottom of $|0\ldots0\ra_{f}$ we denote the basis in which the vector is considered. $\beta$ is the normalization constant for the density matrix given by $Tr(\rho)=1$, which can be expressed in the  Liouville-Fock space as:
\be {}_f\la I | \rho_{\infty}\ra_{f}=1.\nonumber\ee
The left vacuum of the Liouvillian in the diagonal basis~(\ref{Ldiag}) is vacuum of creation operators 
\be {}_f\la I |=\alpha {}_f\la 0\ldots0|, \alpha\beta=1.\nonumber\ee
The expectation value of the operator $A$ is expressed in the new basis as
\be \la A \ra = {}_f\la I | A_f | \rho_{\infty}\ra_{f}.\ee


\subsection{Full-counting statistics}

Any mesoscopic system experiences quantum fluctuations.~\cite{Bla} The current is not a strict constant depending on time. 
It fluctuates. It is a consequence of the quantization of the electron charge. 
The fluctuation of the current is called shot noise. 
Shot noise contains more information about the system then the current itself. 
For example, it can show if electrons in the system are interacting or not, 
or it reflects the presence of disorder~\cite{Naz97}. 

In principle, one can determine the whole statistics of the electrons transversing the system by introducing 
counting field $\xi$~\cite{Lev}. The characteristic function $F(\xi)$ of the distribution  
\be e^{-\mathcal{F}(\xi)}=\sum_n P(n)e^{i n \xi} \nonumber \ee
gives information about all its cumulants:
\be C_k=-(-i\partial_{\xi})^k \mathcal{F}(\xi)|_{\xi=0}.\ee
The first cumulant gives the current and the second the shot noise:
\be I=\frac{e C_1}{\tau}, S= \frac{2e^2 C_2}{\tau}, \nonumber\ee
where $\tau$ is time, $e$ is the electron charge.
The ratio of the current to the noise is the Fano factor:
\be F=\frac{C_2}{C_1} \label{fano}.\ee
One can also introduce the ratio of the higher cumulants to the first one as a characteristic of the distribution:
\be F_i=\frac{C_i}{C_1}.\ee
The are called generalized Fano factors. 

Let us follow Ref.~\onlinecite{Groth,Carlo06} and introduce the counting field $\xi$ in the Lindblad equation. 
The coupling to the reservoirs brings one electron into or out of the system. 
The counting field at the site $\mu$, where we count the number of electrons having hopped into or out of the system,
enter the Liouvillian in the combination: $$\sum_{in,out} e^{i\xi\sigma_{\mu}^{(in/out)}}L^{(in/out)}_\mu \rho_{\xi}(t)L_\mu^{\dg(in/out)},$$ with 
$\sigma^{(in)}=+1$, $\sigma^{(out)}=-1$.
For example, for the counting electrons hopping to the bath in a one-dimensional geometry one changes the matrix elements $(n+1,n)$ and $(n,2n)$ in the matrix $M$~(\ref{m}) as $-2 \Gamma_1\longrightarrow -2 \Gamma_1 e^{i\xi_1}$ (hopping into the system) and $2\Gamma_2 \longrightarrow 2 \Gamma_2 e^{-i\xi_2}$ (hopping out of the system). 
                        
There is no set of eigenvalues $\lambda_i$ of the matrix $M_{\xi}$ obeying ~(\ref{sumlambda}) anymore. Let us define
\be \delta\lambda[\xi]= \sum_{\lambda_j:\Im\lambda_j>0}\lambda_j - i \Gamma_1 - i\Gamma_2-\sum_{k}U_k.
\ee
The generating function is
\be F[\xi]=i \tau \delta\lambda[\xi],\ee
where $\tau$ is time. 

\subsubsection{Cumulants for the two-site system}
\label{subsubsec:cums}

\begin{figure}[ht!]
\begin{center}
A\includegraphics[width=0.75\linewidth]{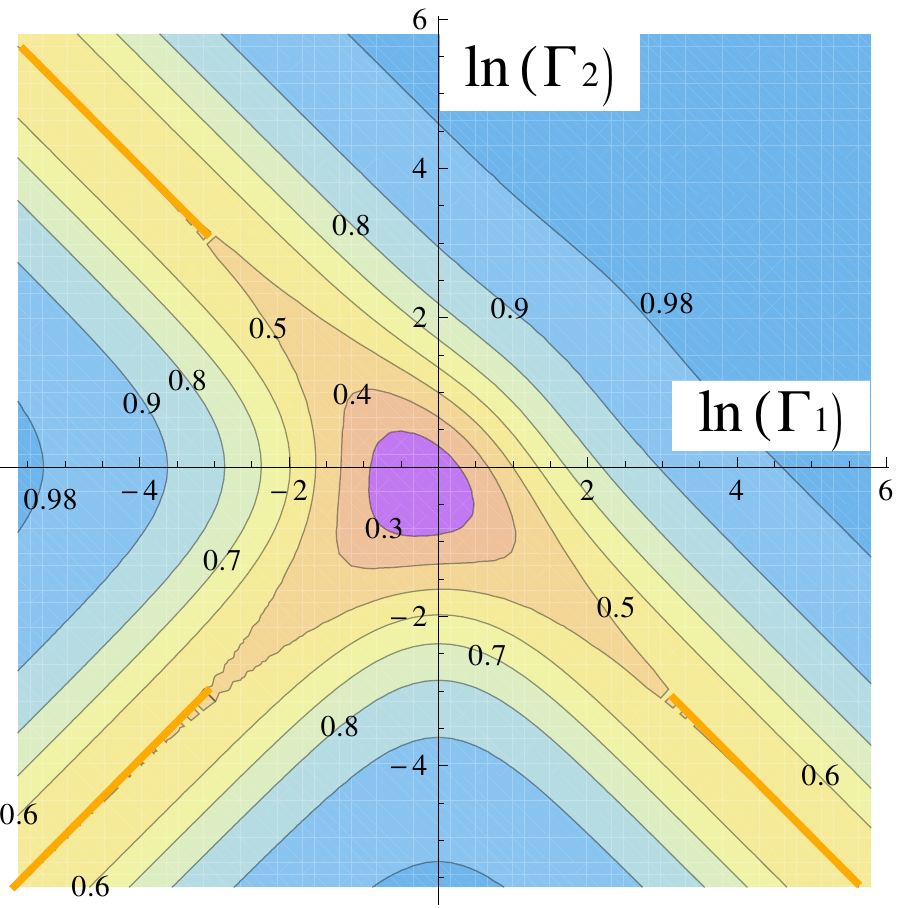}
B\includegraphics[width=0.75\linewidth]{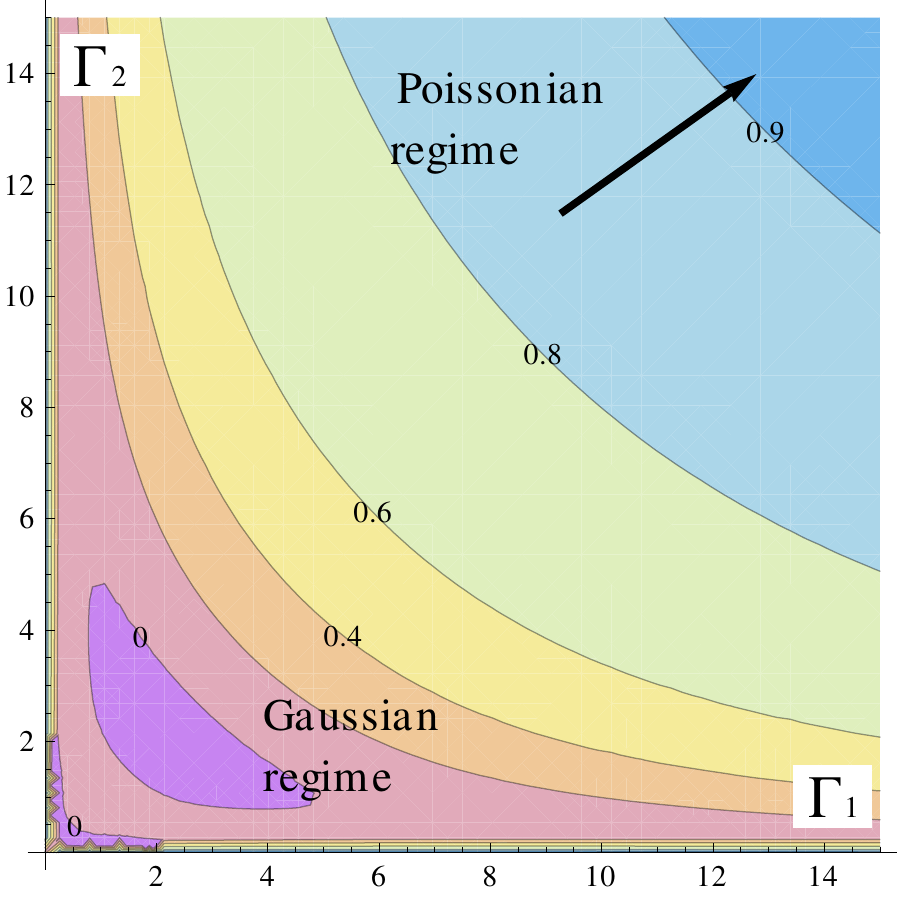}
\caption{ \label{Cums2}
A: The Fano factor for the two-site system depending on $(\ln \Gamma_1,\ln\Gamma_2)$. B:  The second generalized Fano factor depending on the hopping on the both ends $(\Gamma_1,\Gamma_2)$. We clearly see a region where the third cumulant is around zero. Later on, in Fig.~\ref{Cums1}A, we will show that all higher cumulants are close to zero as well. We make all the plots depending on the dimensionless $\Gamma_i$, measured in the units of $t$.}
\end{center}
\end{figure}

\begin{figure}[ht!]
\begin{center}
\includegraphics[width=0.8\linewidth]{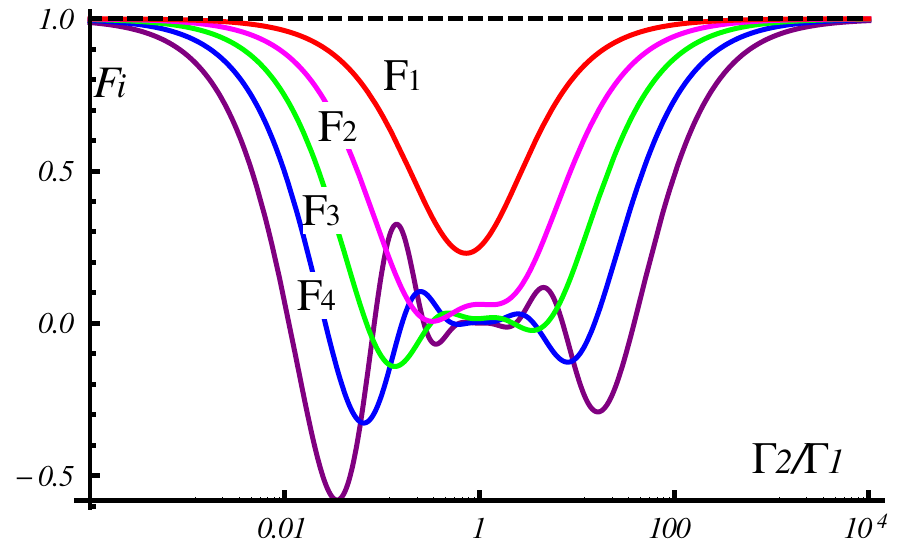}
\caption{ \label{Cums1}
The Fano factor and generalized Fano factors depending on the ration $\Gamma_2/\Gamma_1$ for the two-site system. For the large and small rations of the hybridizations all Fano factors tend to one. There is a region around $\Gamma_i ~ 1$ where the the higher cumulants starting from  the third one are close to zero. They signalize that the  statistics is close to Gaussian.}
\end{center}
\end{figure}

Let us illustrate the above formalism on the example of the two-site system. We can write down the generating function explicitly and hence analyze all cumulants:
\bea F_2[\xi] = \tau (-\Gamma_1 - \Gamma_2 + \sqrt{-2t^2 + \Gamma_1^2 + \Gamma_2^2 + 
  2 g(\xi) } ), \label{generating2} \\
 g(\xi)=\sqrt{4 e^{i \xi} t^2 \Gamma_1 \Gamma_2 + (-t^2 + \Gamma_1 \Gamma_2)^2}, \nonumber
 \eea
where $\xi$ is a difference between the counting fields on the right and on the left, therefore it is responsible for the current conservation.
The first cumulant has a simple form
\be C_1= \tau \frac{2\Gamma_{2}\Gamma_{1} t^2}{(t^2 + \Gamma_{1} \Gamma_{2})(\Gamma_{2}+\Gamma_{1})} \label{cum1}\ee
and $C_1/\tau$ coincides with the local current~(\ref{current}) computed in NESS using $\rho_{\infty}$ (for more discussion see the next Section). 
The second cumulant is:
\be \label{second} C_2 =\tau
\frac{2 t^2\Gamma_2 \Gamma_1 (\Gamma_1^2 \Gamma_2^2 (\Gamma_1 + \Gamma_2)^2 - 
 2 \Gamma_1^2 \Gamma_2^2 t^2 + (\Gamma_1^2 + \Gamma_2^2) t^4)}{(\Gamma_2 + \Gamma_1)^3 (t^2 + \Gamma_2 \Gamma_1)^3}.
\ee
We plot the dependence of the Fano factor on the logarithms of the couplings $(\ln \Gamma_1,\ln\Gamma_2)$  at Fig.~\ref{Cums2}A (we make all the plots depending on the dimmensionless $\Gamma_i$, measured in the units of $t$). 
The higher cumulants for the transport process can be expressed as fractions of the polynomials of $\Gamma_1$ and $\Gamma_2$, which can be directly seen by differentiating the generating function~(\ref{generating2}).

Can we characterize the statistics of the NESS? 
For the hopping through a single-site the statistics is Poissonian for low and high hybridizations and sub-Poissonian for intermediate ones~\cite{Groth}. 
The situation becomes different for the current through the two-site system. For the intermediate hybridization the generalized Fano factors of the distribution
are closer to zero, signalizing that the statistics is close to Gaussian  for this hybridization strength. To illustrate this we plot the dependence of the generalized Fano factors on $\Gamma_1$ for fixed $\Gamma_2$, Fig.~\ref{Cums1}. The second generalized Fano factor is shown at Fig.~\ref{Cums2}B reflecting some region where the cumulant is close to zero. For the large and small ratios of the hybridizations the statistics is still Poissionian, Fig.~\ref{Cums1}. 
The physical reason for the smallness of the cumulants for intermediate
couplings is that the electons need some time to hop into and out of the
system. For the large or small ratios of the hybridizations $\Gamma_1$ and
$\Gamma_2$ the statistics is largely determined by the hopping at one end
only and becomes Poissionian.

\section{Uniform system}
\begin{figure*}
\begin{center}
A\includegraphics[width=0.45\linewidth]{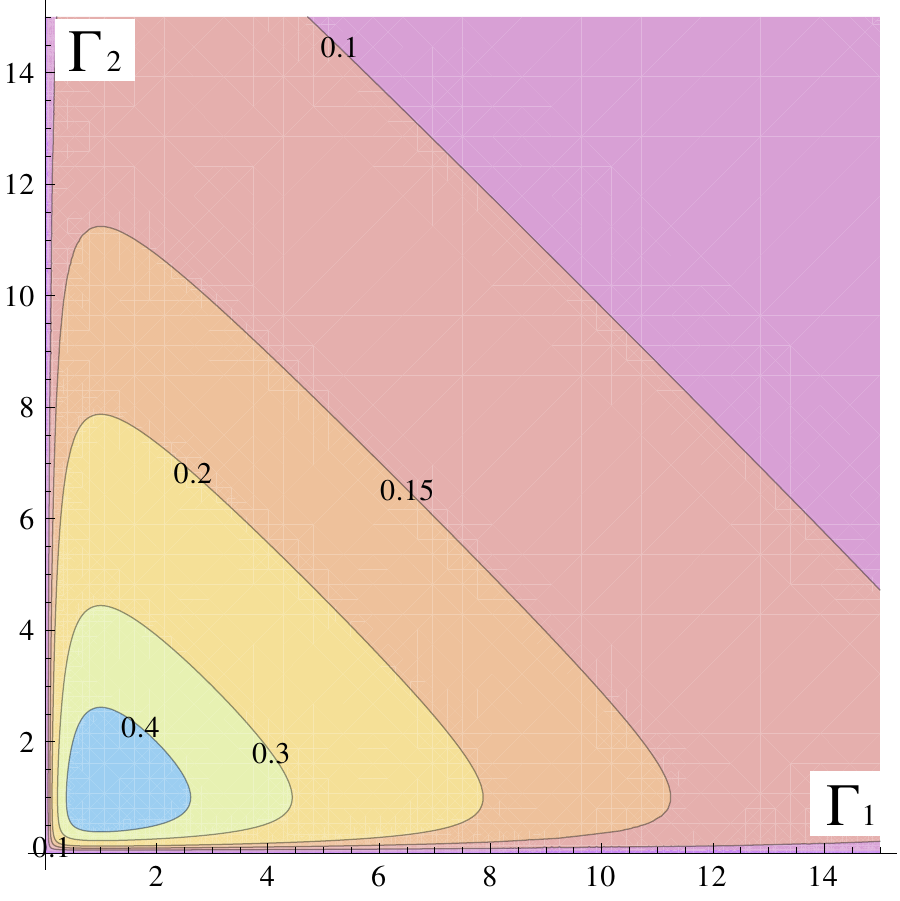}
B\includegraphics[width=0.45\linewidth]{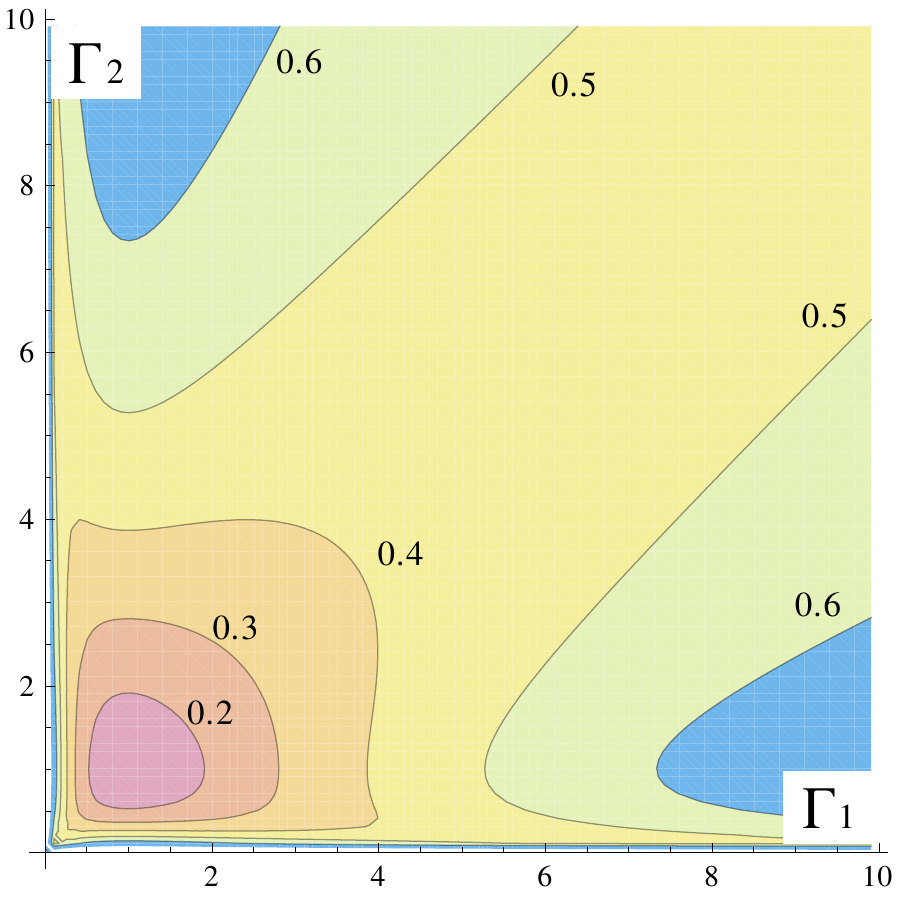}
C\includegraphics[width=0.45\linewidth]{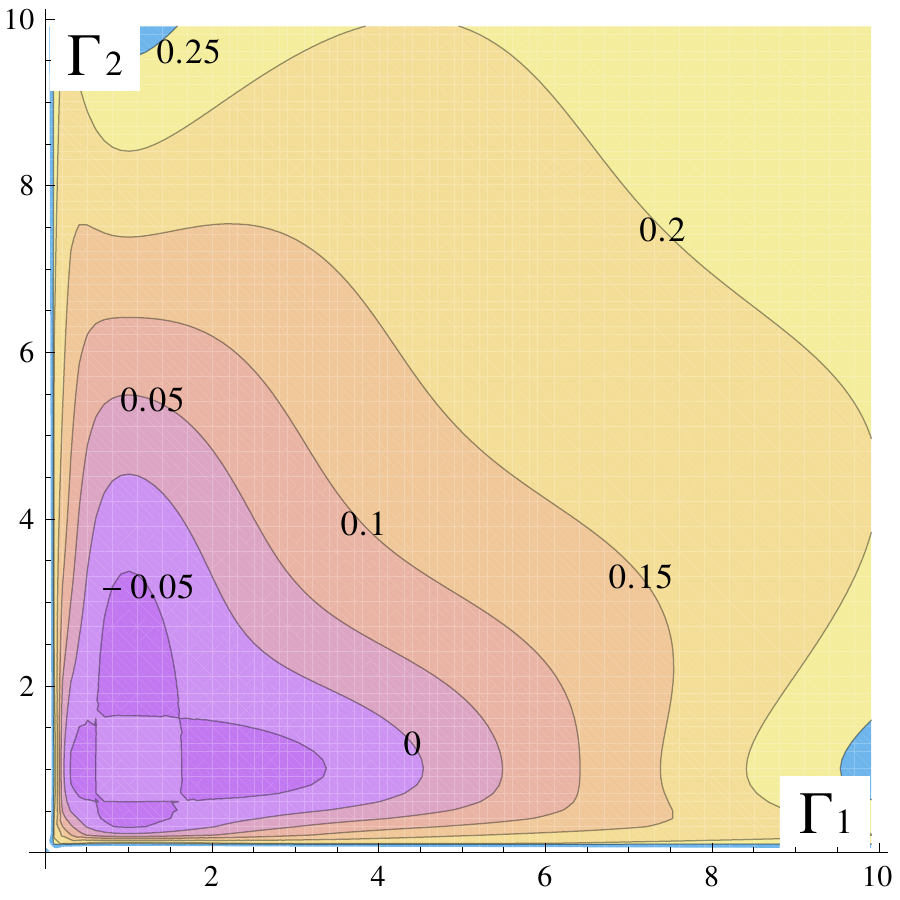}
D\includegraphics[width=0.45\linewidth]{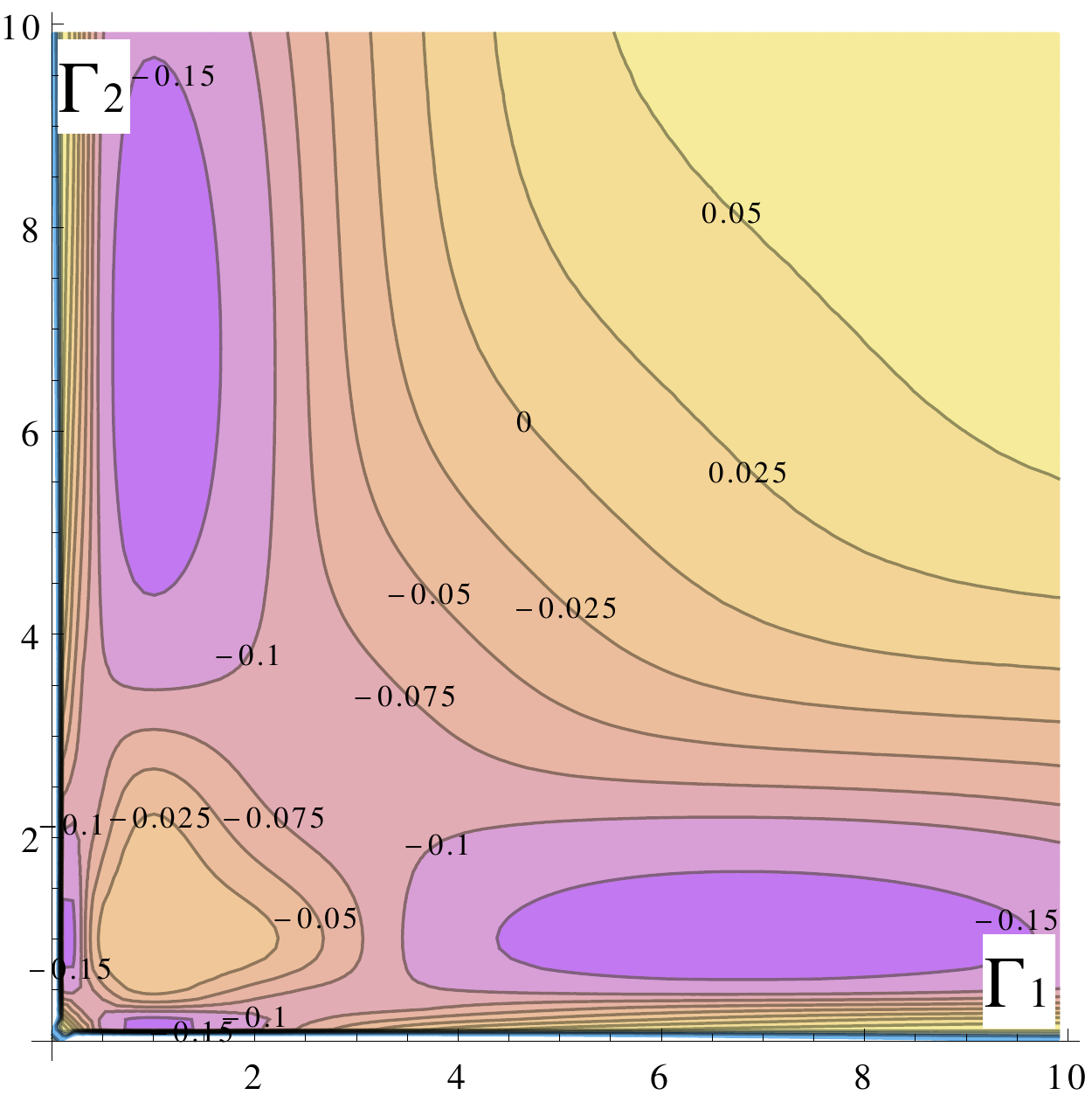}
\caption{ \label{fanos}
Dependence of the current (A), Fano factor (B), second (C) and third (D) generalized Fano factor on hybridizations with the bath for system sizes large enough that these quantities do not depend on the system size, i. e. larger or equal to 2, 3, 4, 5 sites, respectively.}
\end{center}
\end{figure*}

Let us use the formalism described in the previous section to compute the current and its cumulants for the long uniform system.

The calculation of the current can be performed in two ways: by
averaging  the local current operator over NESS
\be \hat{j_k} =-i t (a_{k}^\dg a_{k+1}+a_{k+1}^\dg a_k) \label{current}\ee
and by differentiating the generating function. The answer is the same, we also notice that the current is uniform along the system in NESS. The same expression for the current can be obtained using the Green function formalism and the Meir-Wingreen expression for the current in the limit of the infinite bias voltage in the leads (see Appendix~\ref{app:Green}). 
The situation with the higher cumulants is different: the local correlation function for the current at different points is different from the results got via the differentiation of the generating function. The noise calculated by the Green function approach is different from the noise obtained from the counting statistics for the Lindblad equation. The Green function approach takes into account the interference in the leads automatically, while the counting statistics derived from the memory-less Lindblad equation does not know anything about the interference in the leads. 
We suggest that the current is the same in all three approaches as it represents a fundamental property of the charge conservation conservation, while the current noise characterizes the statistics of the charge transfer and it is different  inside the system where the charge transfer is quantum-mechanical process, while the transfer from the system to the bath is classical in the Lindblad formalism.

\subsection{Current}
The current through the uniform system does not depend on the system size when the number of sites in the lattice is greater
than 2, and is given by the expression~(\ref{cum1}). For the two-site system we obtain this expression analytically both by averaging the current operator (\ref{current}) over the NESS density matrix (formalism of the Sections~\ref{subsec:tilde}-\ref{subsec:newbasis}) and from the generating function (Section~\ref{subsubsec:cums}).
We expect this correspondence between the results as the current inside the system and the current out of the system is the same in the absence of external sources due to the charge conservation. 

The current through the electronic system connected to the non-interacting reservoirs can be described by Meir-Wingreen formula,~\cite{MW} see Appendix~\ref{app:Green}. For the case of  infinite bias voltage in the leads the current through the system is given again by~(\ref{cum1}). 
For the case of interacting reservoirs the value of the current differs from the result~(\ref{cum1}), see Appendix~\ref{app:interacting}. 
One expects both of these results as the derivation of the Lindblad equation for the case of infinite bias voltage from the wave function formalism~\cite{Gurv} is valid only in the case of the non-interacting reservoirs. On the other hand, there is a special master equation which takes into account only the sequential tunneling from the bath to the leads for the case of Luttinger liquid baths~\cite{LLSTA}. 

Let us discuss the physical nature of the dependence of the current on the hybridization with the bath, Fig.~\ref{fanos}A.
The current increases with increasing the couplings to the bath $\Gamma_1$ and $\Gamma_2$ for $\Gamma_1, \Gamma_2 \lesssim 1$ because the probability of hops is increased. For larger couplings the current decreases. It is the signature of the quantum Zeno effect: the quantum system is constantly being measured by the coupling to the classical bath, this measurement localizes the states at the ends of the chain, hence the current decreases. 

\subsection{Noise}
We have already obtained the noise for the two-site system, (\ref{second}) and Fig.~\ref{Cums2}A. 
For the system of length longer than three sites the Fano factor does not depend on the system size:
\be F= \frac{\Gamma_1^4\Gamma_2^2+t^4(\Gamma_1^2+\Gamma_2^2)-2t^2\Gamma_1^2\Gamma_2^2+\Gamma_1^2\Gamma_2^4}{(\Gamma_1+\Gamma_2)^2(\Gamma_1\Gamma_2+t^2)^2}.\ee
Its functional dependence on the couplings to the leads is shown on Fig.~\ref{fanos}B. 
The noise increases with increasing the couplings to the bath.

What local operator evaluated in NESS can we write in correspondence to the noise? 
We can consider the following discretized version of $\la \hat{j}^2 \ra - \la \hat{j} \ra^2$:
\be \label{OpNoise}\hat{S}_k=tr((\hat{j}_k\hat{j}_{k+1}+\hat{j}_{k+1}\hat{j}_{k}) \rho_{\infty})/2 -( tr(\hat{j}_k\rho_{\infty}))^2.\ee
This operator also does not depend on the system size for the systems larger than three sites.  
The second term does not depend on the size of the system as we discussed earlier. 
The operator expression corresponding to the first term is $a^\dg_{k}a_{k+2}+a^\dg_{k+2}a_{k}$ and its the average over NESS $\la \hat{j} \ra^2$.
It is possible to construct the analogues of the higher cumulants independent of the system size using the same type of the discretization of the averages  of $\hat{j}^n$ and show that they depend only on $\la \hat{j}\ra$. 

\begin{figure}[ht!]
\begin{center}
\includegraphics[width=0.85\linewidth]{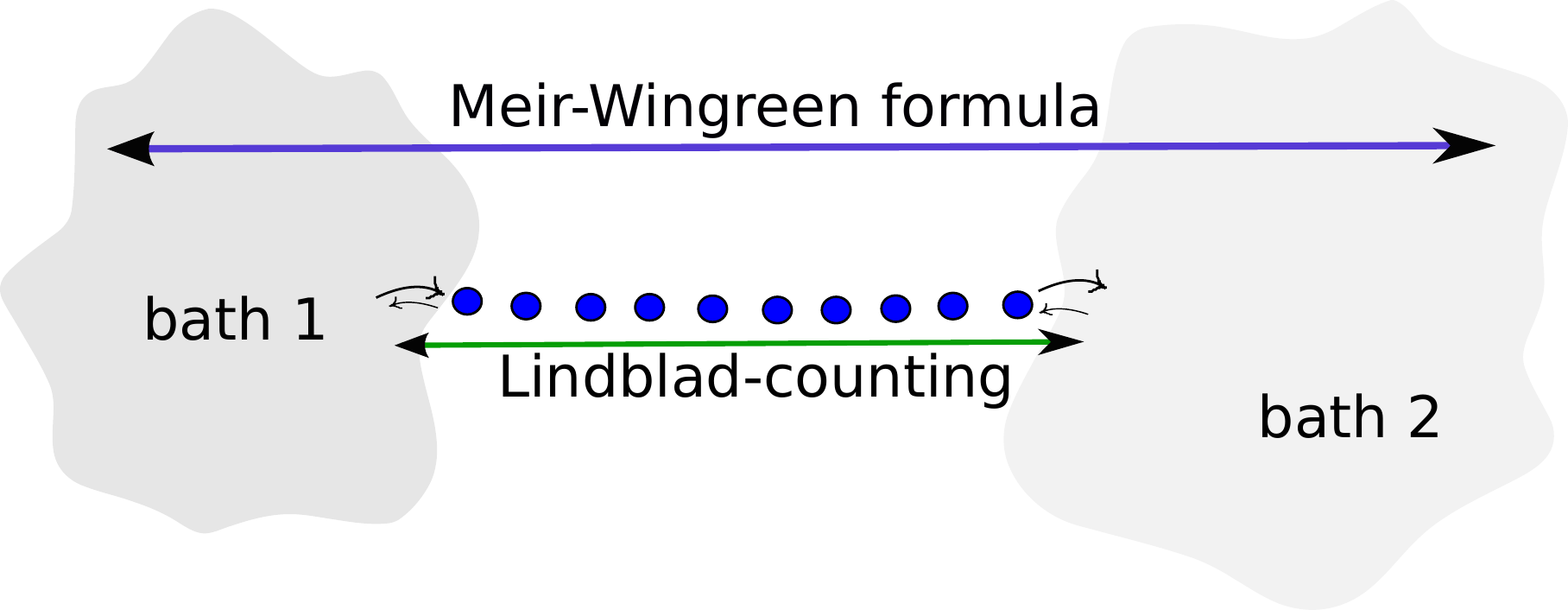}
\caption{ \label{setup}
Schematic representation of the charge transport statistics measurement. 
The counting statistics of the electrons from the Lindblad equation corresponds to the measurement of the charge statistics right at the connection of the quantum system to the reservoirs, while the Meir-Wingreen-type of the expression for the current and the noise takes into account also the interference in the leads and corresponds to the measurement in the reservoirs far away from the system.}
\end{center}
\end{figure}

\begin{figure}[ht!]
\begin{center}
A\includegraphics[width=0.85\linewidth]{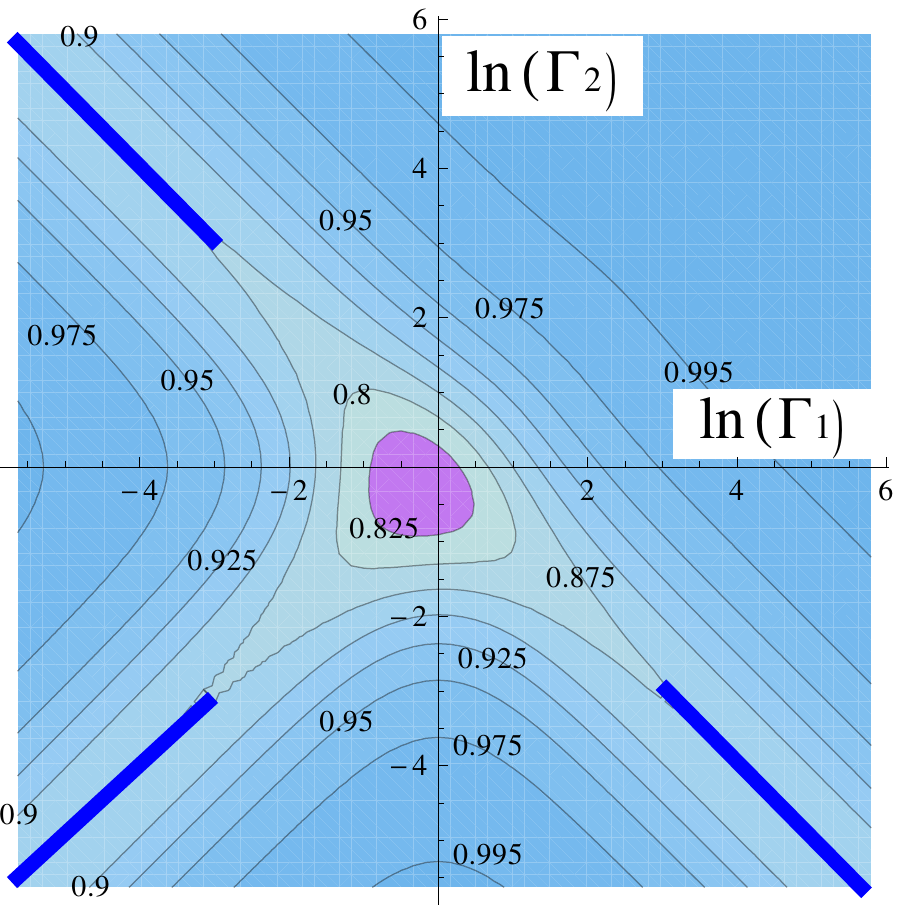}
B\includegraphics[width=0.85\linewidth]{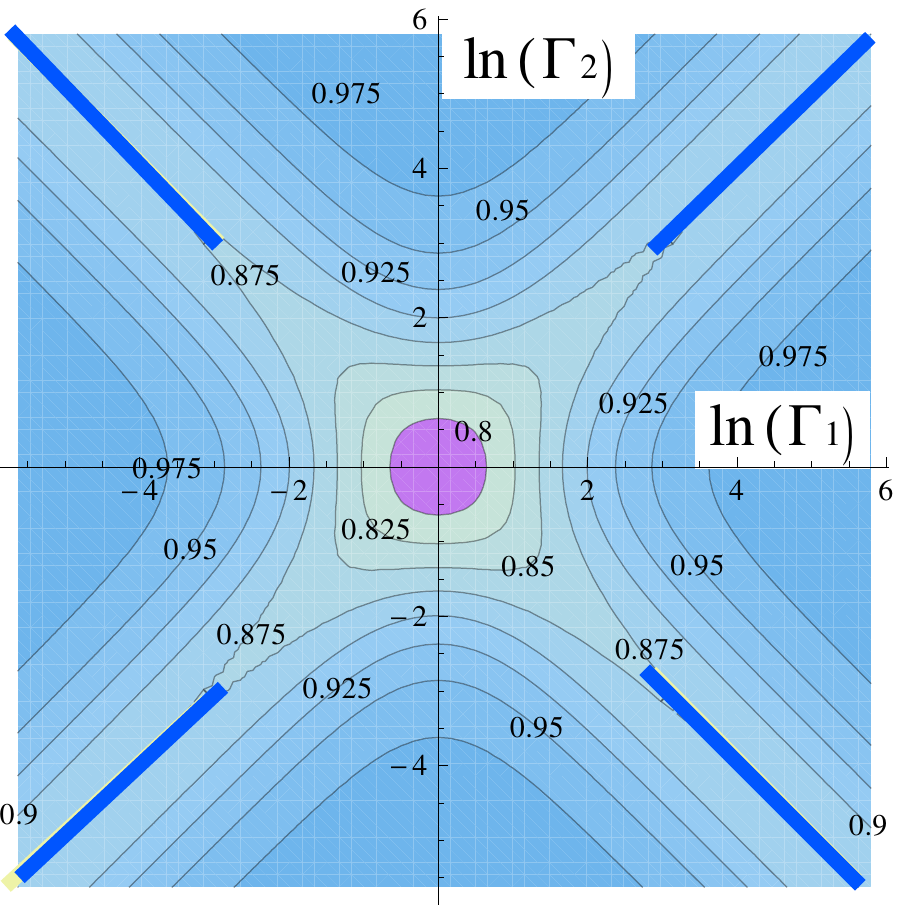}
\caption{ \label{curr}
The Fano factor calculated by the Green function formalism for the two-site systems (A) and three-site systems (B). 
Comparing the contours in these two plots to the contours in Figs.~\ref{Cums2}B and~\ref{logfanos}A, one can see that  the linear relation~(\ref{ratio}) between $F_{MW}$ and $F_L$ is valid.}
\end{center}
\end{figure}

Let us compare the noise obtained using the counting field in the Lindblad equation and by the Green function technique, see Appendix~\ref{app:Green}.
For the one-site system both the noise and the current are the same by both methods. 
For the two-site system we derive an analytical expression by both methods and see that the Fano factors are connected by the simple linear relation:
\be F_{MW}=\frac{F_{L}+3}{4},\label{ratio}\ee
where we denote by the index $MW$ the answer obtained by evaluating the Meir-Wingreen expression, and the subscript $L$ is for the electron statistics got by counting electrons in the Lindblad formalism; elsewhere in the paper we omit the subscripts $L$ and $MW$ as we are mostly discussing the Lindblad approach.
For larger system sizes the analytical comparison becomes more complicated and we examine the integrals in Meir-Wingreen formalism numerically. We find that the current does not depend on the system size for sizes greater than size $2$, nor does  
the noise depend on system sizes greater than $3$. Crucially, {\it the linear relation~(\ref{ratio}) between the Fano factors in the two approaches remains fixed for the systems with more than 1 site}.
There is no a linear relation between the further generalized Fano factors. 
The usual Fano factor for non-interacting electrons is always smaller then 1. 
Therefore,{\it $F_{MW}$ is always larger then $F_L$}. 
What does this mean?
The transport properties calculated in the Green function formalism take into account the interference in the reservoirs, 
while the Lindblad formalism describes a quantum system coupled to a completely memory-less bath. 
$F_{MW}$ accounts not only for the noise of the system itself, but also for the noise during the propagation in the reservoirs. 
We suggest that one can distinguish between $F_{MW}$ and $F_L$ in experiment by putting the contacts used to measure the statistics of the charge transport at different positions:
for $F_L$ one should place contacts as close to the system as possible, 
while for $F_{MW}$ the contacts should be made far from the system itself, inside the reservoir (Fig.~\ref{setup}).

\subsection{Hierarchy of the cumulants for the discretized model}
\begin{figure}[tb!]
\begin{center}
A\includegraphics[width=0.85\linewidth]{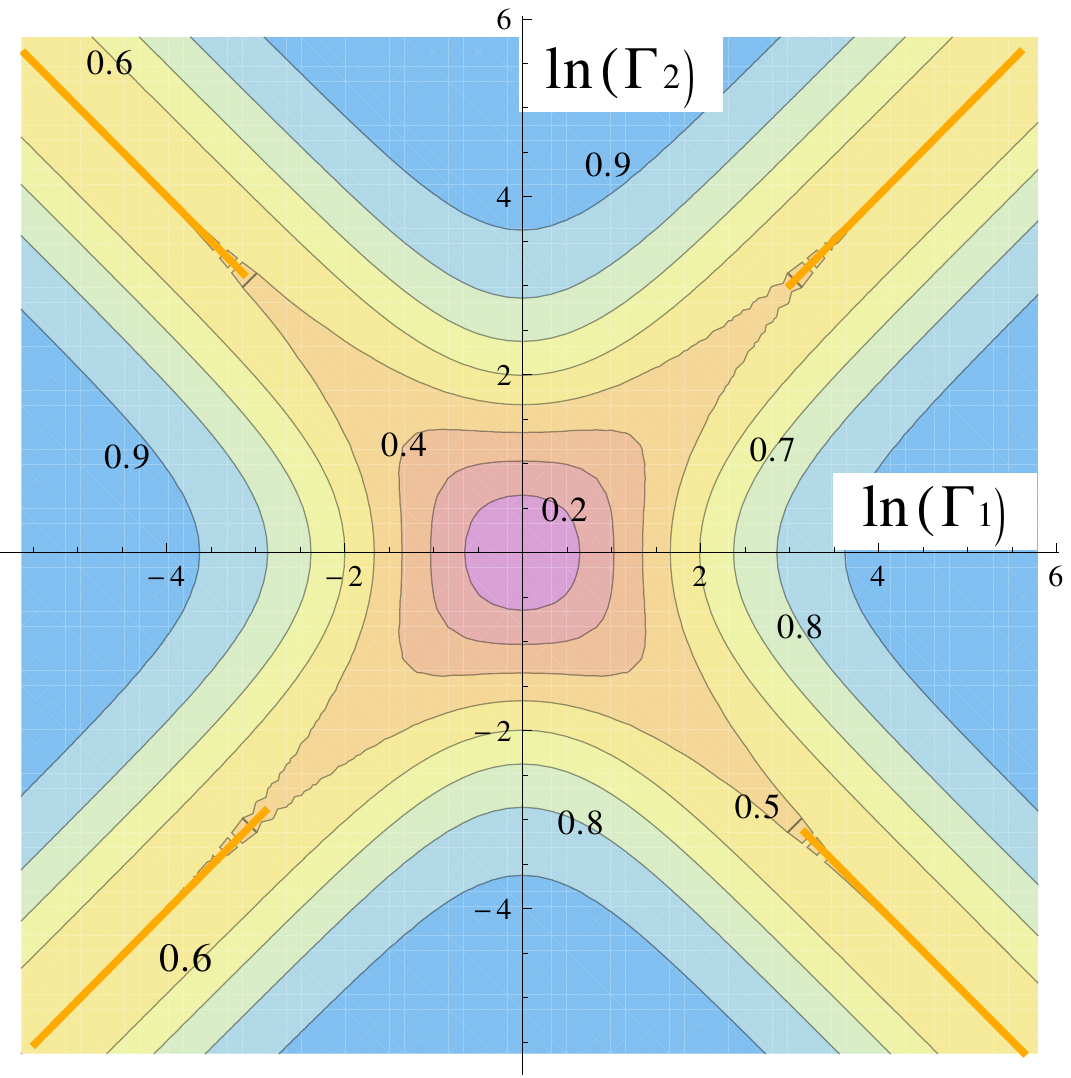}
B\includegraphics[width=0.85\linewidth]{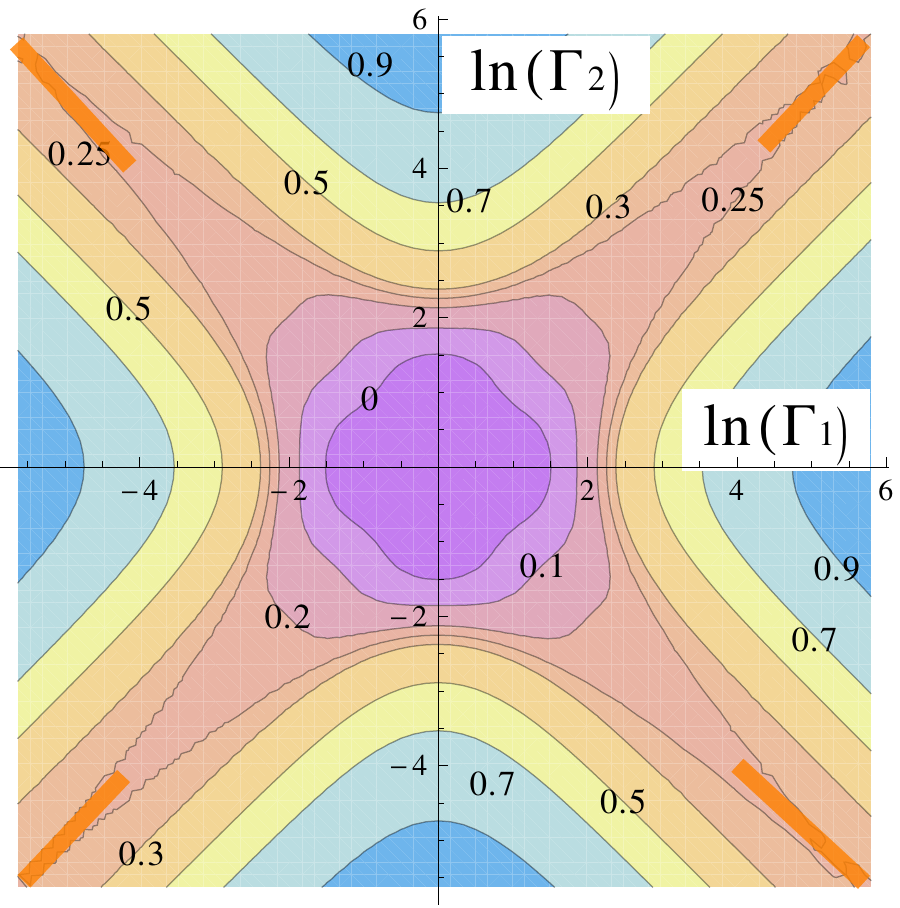}
\caption{ \label{logfanos}  Dependence Fano factor (A) and the second generalized Fano factor (B)
on the logarithms of the hybridizations, $(\ln \Gamma_1,\ln\Gamma_2)$. 
In the plot we clearly see that $F_i(x,y)=F_i(y,x)=F_i(1/x,1/y)$.
The asymptotic behavior for the $\Gamma_1=\Gamma_2 \mapsto \infty$ is denoted by the thick line. The minimum of the generalized Fano factors is reached for $\Gamma_1=\Gamma_2=1$.}
\end{center}
\end{figure}

According to our numerical investigation we can formulate the following conclusion: 
{\it The n-th cumulant of the charge transport for the uniform tight-binding chain
does not depend on the chain length for chains longer than 
$n+1$ site.} 
In this limit, the cumulants as functions of the couplings are symmetric with respect to change $\Gamma_i\mapsto 1/\Gamma_i$, Fig.~\ref{logfanos}. For systems shorter than $k+1$, the cumulant of order $k$ does not have this property, see for example Fig.~\ref{Cums2}B (for the one-site system it is not true: all generalized Fano factor posses this property).


We also find the following asymptotic behavior of the generalized Fano factors for $\Gamma_1=\Gamma_2\longrightarrow \infty$, see Fig.~\ref{logfanos}: 
\bea  F&\longrightarrow& 2^{-1}, \nonumber\\ 
F_2&\longrightarrow& 2^{-2}, \nonumber\\
F_3&\longrightarrow& 2^{-3}. \nonumber
\eea
This leads us to the straightforward generalization:
\be F_n \longrightarrow 2^{-n}\ee
and gives us full knowledge about the statistics of the random process  for $\Gamma_1=\Gamma_2\longrightarrow \infty$. 

We can also consider another limit: $\Gamma_1\ll \Gamma_2\longrightarrow \infty$ or $\Gamma_2\ll \Gamma_1\longrightarrow \infty$. 
In this case all cumulants are equal and the Fano factors $F_i\longrightarrow 1$. 
This signalizes the Poissonian nature of the charge transport determined by the lowest of the couplings, $\min(\Gamma_1,\Gamma_2)$, which limits the magnitude of the current. 

\section{Conclusions}
In the present work we have investigated the full-counting statistics of an open non-interacting fermionic system 
coupled to the classical Markovian bath. Such systems are described by the Lindblad equation. We generalize the 'super-operator' formalism of Ref.~\onlinecite{Kosov} 
by including the counting field. The generating function of the statistics is connected to the trace of the matrix describing the tight-binding structure of the chain and the structure of the dissipative couplings in the presence of the counting field. 

We derive an analytical expression for the generating function of the two-site system. The generating function for the one-site system has been derived and investigated earlier in Ref.~\onlinecite{Groth}. To the best of our knowledge, analytical expression for the two-site system has not been discussed before. 
For systems of the longer than three sites we perform numerical investigation. 
In the uniform one-dimensional system we find that the generalized Fano factors of order $k$ are independent of the system size for the system size starting from $k+1$, for example, the current does not depend on the system size for lengths larger than two, the Fano factor -- for systems longer than three, the second generalized Fano factor -- longer than four, etc. It means that for the infinitely long system the current statistics does not depend on the system length.
There is a duality in the behaviour of the cumulants for the large and small hybridization strengths to the reservoirs. 
The local current in NESS is constant along the chain and the same as obtained from the counting statistics approach due to the current conservation in the system in the absence of external sources and drains. 


On one hand, the Lindblad equation for the system connected to non-interacting reservoirs can be derived from the evolution of the wave function~\cite{Gurv} under the assumption of the infinite bias voltage in the leads. 
On the other hand, the non-interacting fermionic system coupled to non-interacting fermionic bath is a simple system and can be solved exactly purely analytically using for example the Green function method.
We compare the first two cumulants of the charge transport statistics given by these two approaches.
The current in both approaches is the same and does not depend on the system size, which is again a consequence of the charge conservation. 
The noise given by the Green function method is always larger than as obtained by the Lindblad counting. One could guess this result: the Lindblad equation does not take into account memory effects in the bath, while the Green function calculation takes into account the interference in the reservoirs and hence memory effects. 
We suggest that experimentally the counting statistics is different depending on the position of the measuring contacts: 
the contacts close to the system will measure the Lindbladian statistics, while the contacts deep in the reservoirs will confirm the statistics given by the scattering type/Green function approach.

Our result about the generating function for the full-counting statistics for an open quantum system of non-interacting fermions is closely related to the result of preprint~\onlinecite{Buca}, which we noticed at the late stage of the preparation of our manuscript. 
\section*{Acknowledgments}
This work was supported through SFB 1073 of the Deutsche Forschungsgemeinschaft (DFG).
We are grateful for helpful discussions to Mihailo \u{C}ubrovi\'{c} and Ebad Kamil.

\appendix

\section{Green function calculation}
\label{app:Green}

The current through the system connected to the non-interacting leads can be determined via 
the Meir-Wingreen formula~\cite{MW}. This expression involves the Green function 
of the system $\mathbf{G}$ renormalized by the interaction with the leads.
The current through a non-interacting system is~\cite{MW}:
\be I_{MW} = \frac{e}{\pi} \int d\omega (f_1(\omega) - f_2(\omega) ) tr [\mathbf{G^a \Gamma^{(2)} G^r \Gamma^{(1)}}] \label{MW}.\ee
What does~(\ref{MW}) mean? It gives the transmission coefficient at fixed energy  $t t^\dg = \mathbf{G^a \Gamma^{(2)} G^r \Gamma^{(1)}}$. Using it we can determine also the noise:
\be S_{MW} = \frac{e}{\pi} \int d\omega (f_1(\omega) - f_2(\omega) ) tr [t t^\dg (1-t t^\dg)].\label{noise}\ee

$\mathbf{G}$ is given by the Dyson equation:
\be \mathbf{G}=(\mathbf{G}_0^{-1}-\mathbf{\Sigma})^{-1}. \label{Dyson}\ee
The Green function $\mathbf{G}_0$ of the system uncoupled to the environment is determined from the solution of the Schroedinger equation:
\be \mathbf{G}_0(\omega)=\sum_i \frac{\psi_i \psi_i^\dg}{\omega-E_i}, H\psi_i=E_i\psi_i.\label{bare}\ee

The self energy $\mathbf{\Sigma}$ accounts for the tunneling from the first/last site of the chain to the reservoir, propagation in the reservoir, tunneling back to the first/last site. Let us notice that the reservoirs are non-interacting, meaning that the momentum is conserved during the propagation.
Therefore, 
\be \Sigma^{(1/2),r/a}(\omega)= |V_{1/2}|^2\int \frac{d^{d}k }{\omega-(k^2/2m-\mu_{1/2})\pm i0} \label{sigma}\ee
(here we assumed that couplings to different $k$ modes are the same; the reservoirs are $d$- dimensional). We also assume the chemical potential in the leads to be very large, then the leading part of the integration in (\ref{sigma}) comes from the region around the Fermi energy $\mu_{1/2}$. We linearize the denominator and get from the Sokhotsky formula: 
\be \Sigma^{(1/2),r/a}(\omega)=\mp 2i \pi \frac{\nu}{(2\pi \hbar)^d} |V_{1/2}|^2,\ee
$\nu$ is a density of states at the Fermi-energy, 2 comes from the contribution around $\pm k_F$. Let us in the following put $2\pi\hbar$ to 1. 
The self-energy is a constant in the limit of large chemical potential in the leads.
The matrix structure of the self-energy is ${\mathbf \Sigma^{(1)}}_{1,1}=\Sigma^{(1)}$,
${\mathbf \Sigma^{(2)}}_{N,N}=\Sigma^{(2)}$, the other elements of the both matrices are zeros. 

Thw couplings $\mathbf{\Gamma^{(1/2)}}$ are defined by  $\mathbf{\Gamma^{(1/2)}}=i\mathbf{\Sigma^{(1/2),r}}$. 
For any tight-binding Hamiltonian one can derive the bare Green function of the system $\mathbf{G}_0(\omega)$, ~(\ref{bare}), and evaluate the current and the noise according to~(\ref{MW}) and (\ref{noise}). The results of such evaluation and comparison with the Lindblad formalism are discussed in the main text of the paper. 




\section{Transport properties for the case of interacting leads}
\label{app:interacting}

\begin{figure}[ht!]
\begin{center}
A\includegraphics[width=0.85\linewidth]{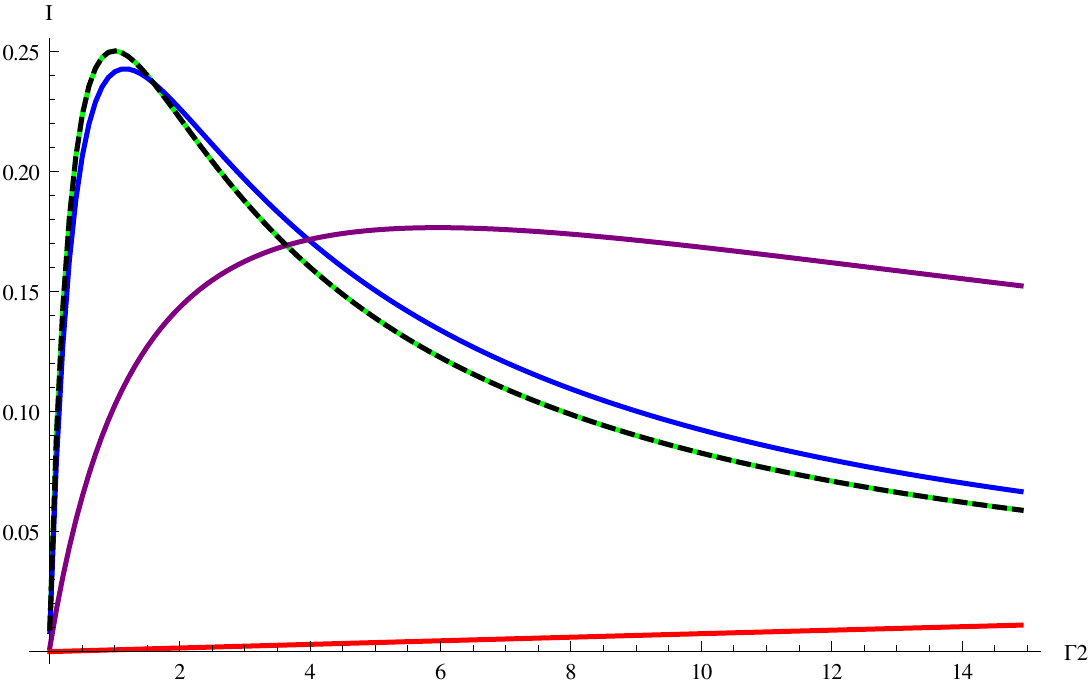}
B\includegraphics[width=0.85\linewidth]{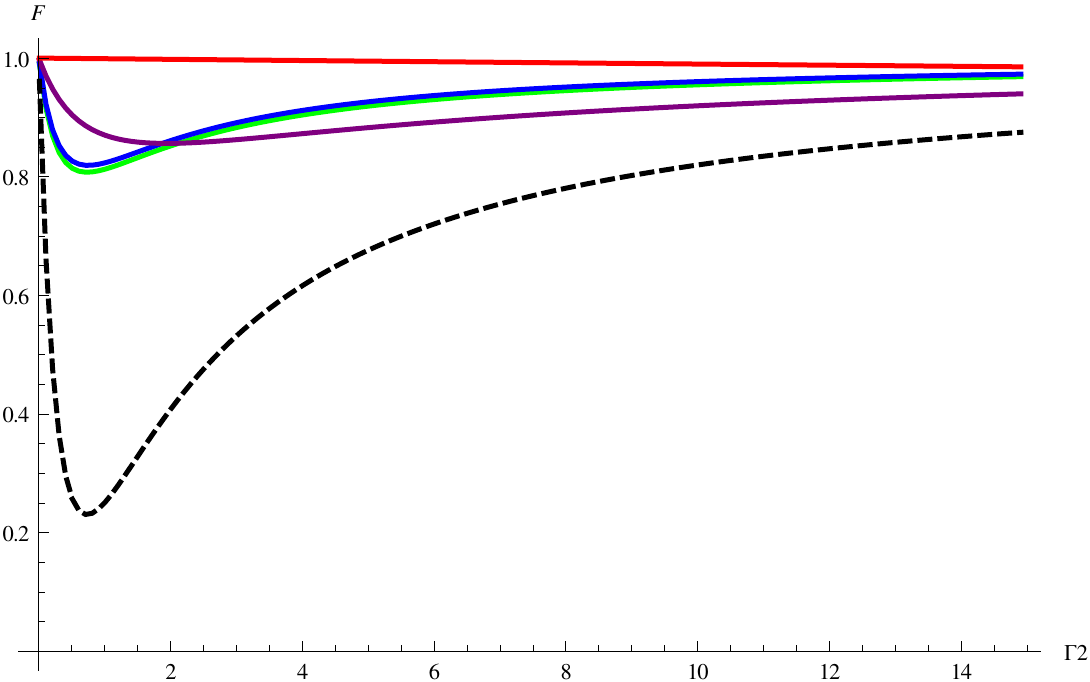}
\caption{ \label{FigLL}
A: The current, B: The Fano factor for different interaction strengths in the Luttinger liquid reservoirs (solid lines: green $g=1.$ -- non-interacting case, blue $g=0.8$ -- weak interaction, purple $g=0.4$, red $g=0.2$). 
Dashed black line at plot B represents the noise calculated using the counting statistics from the Lindblad approach. 
Let us notice that the Lindblad type of the noise calculation gives the Fano factor smaller that at any interaction strength $g$. }
\end{center}
\end{figure}

Here we give a brief summary of the calculation of the current and the noise through a non-interacting chain coupled to the Luttinger liquid reservoirs. We follow the references \onlinecite{Ch1, Ch2} where the generalizations of the Meir-Wingreen formulas for the interacting reservoirs are derived. The Green function in the Luttinger liquid reservoir is taken from Ref.~\onlinecite{LLG}.


\subsection{Luttinger liquid reservoirs}

We derive a zero-temperature asymptotic form of the expressions given in reference~\onlinecite{Ch1,Ch2} using the asymptotic expansion of the Gamma-function. 
Most probably, a very close result can be obtained by  the master equation approach for the transport between two Luttinger liquids~\cite{LLG,LLSTA}. We use the Green function approach as 
we do not assume that only sequential tunnelling contributes, unlike the derivation in Ref.~\onlinecite{LLSTA}, which is thus valid for small tunnelling amplitudes only.

The current and the noise through the system connected to the interacting reservoirs are~\cite{Ch1,Ch2}:
\begin{widetext}
\bea I &=& \frac{e^2}{2 \pi } \int d\omega tr (\mathbf{\Gamma}_1 \mathbf{G}^{r}(\omega)\mathbf{\Gamma}_2 \mathbf{G}^{a}(\omega)) [F^{<}_1(\omega)F^{>}_2(\omega)-F^{>}_1(\omega)F^{<}_2(\omega)] \\ 
S &=& \frac{e^2}{2 \pi } \int d\omega tr (\mathbf{\Gamma}_1 \mathbf{G}^{r}(\omega)\mathbf{\Gamma}_2 \mathbf{G}^{a}(\omega))
[F^{<}_1(\omega)F^{>}_2(\omega)+F^{>}_1(\omega)F^{<}_2(\omega)]-\\
&-&\frac{e^2}{2 \pi } \int d\omega tr (\mathbf{\Gamma}_1 \mathbf{G}^{r}(\omega)\mathbf{\Gamma}_2 \mathbf{G}^{a}(\omega)\mathbf{\Gamma}_1 \mathbf{G}^{r}(\omega)\mathbf{\Gamma}_2 \mathbf{G}^{a}(\omega)) [F^{<}_1(\omega)F^{>}_2(\omega)-F^{>}_1(\omega)F^{<}_2(\omega)]^2
\eea
where functions $\mathbf{F}^{</>}$ are analogous to the distribution functions of non-interacting electrons, $f(\omega)$ and $1-f(\omega)$. The expressions for them are calculated from the Green functions of the Luttinger liquid and have the form:
\bea F^{<}_{1,2}&=&\frac{1}{2\pi}e^{-(\omega -\mu_{1,2})/2T} \left(\frac{\pi T}{\Lambda} \right)^{1/g-1}\frac{|\Gamma(\frac{1}{2g} + i \frac{\omega - \mu_{1,2}}{2\pi T}|^2}{\Gamma(1/g)} \\ 
F^{<}_{1,2}&=&\frac{1}{2\pi}e^{(\omega -\mu_{1,2})/2T} \left(\frac{\pi T}{\Lambda} \right)^{1/g-1}\frac{|\Gamma(\frac{1}{2g} + i \frac{\omega - \mu_{1,2}}{2\pi T}|^2}{\Gamma(1/g)}.\eea
The Green function of the non-interacting subsystem is calculated from the Dyson equation~(\ref{Dyson}) with self-energy
\be \mathbf{\Sigma}^{r}(\omega) =-i [\mathbf{\Gamma}_1 (F^{>}_1(\omega)+F^{<}_1(\omega))+\mathbf{\Gamma}_2 (F^{>}_2(\omega)+F^{<}_2(\omega))]/2.\ee
Let us note that for the Luttinger liquid the Green function depends on the high-energy cut-off $\Lambda$.

Now we use the asymptotic expression for the Gamma function
\be \ln \Gamma(z) \sim (z-1/2) \ln z - z + \ln(2\pi)/2+O(1/z), z\rightarrow \infty, |\arg z|<\pi.\ee
We put $z=\frac{1}{2g} + i \frac{\omega - \mu_{1,2}}{2\pi T}$ and obtain:
\bea F^{<}_{1,2}&=& \frac{\theta(\omega -\mu_{1,2}) \left(\frac{\omega - \mu}{\Lambda} \right)^{1/g-1}}{\Gamma(1/g)} \\
F^{>}_{1,2}&=& \frac{\theta(-\omega +\mu_{1,2}) \left(\frac{\omega - \mu}{\Lambda} \right)^{1/g-1}}{\Gamma(1/g)}\eea
Then the current and the noise are 
\bea 
I &=& \frac{e^2}{2 \pi } \int_{\mu_1}^{\mu_2} d\omega tr (\Gamma_1 G^{r}(\omega)\Gamma_2 G^{a}(\omega)) \frac{\left(\frac{\omega-\mu_1}{\Lambda} \right)^{1/g-1}\left(\frac{\omega-\mu_2}{\Lambda} \right)^{1/g-1}}{\Gamma^2 (1/g)} \\ 
S &=& I -\frac{e^2}{2 \pi } \int_{\mu_1}^{\mu_2} d\omega tr (\Gamma_1 G^{r}(\omega)\Gamma_2 G^{a}(\omega)\Gamma_1 G^{r}(\omega)\Gamma_2 G^{a}(\omega)) \frac{\left(\frac{\omega-\mu_1}{\Lambda} \right)^{2/g-2}\left( \frac{\omega-\mu_2}{\Lambda} \right)^{2/g-2}}{\Gamma^4 (1/g)}.
\eea
\end{widetext}

We are interested in the limit of the large bias voltage. We take this limit first by using the chemical potential as the cut-off $\Lambda$. Then we take $\Lambda$ large enough that the result does not depend on it.

The transmission characteristics -- the current and the Fano factor --  for different interaction strengths in the Luttinger liquid are shown in Fig.~\ref{FigLL} for the non-interacting system consisting of two sites.
The counting statistics given by introducing a counting field in the Lindblad equation is different from the statistics with both interacting and non-interacting leads. The counting given by the Lindblad equation always gives the reduced value of the noise.


\end{document}